\documentclass[lettersize,hidelinks,conference]{IEEEtran}

\usepackage{cite}
\usepackage{amsmath,amssymb,amsfonts}
\usepackage{graphicx}
\usepackage{textcomp}
\usepackage[dvipsnames]{xcolor}
\usepackage[acronym,shortcuts]{glossaries}
\usepackage{bm}
\usepackage{caption}
\usepackage{subcaption}
\usepackage{relsize}
\usepackage{soul}
\usepackage{algorithm, algpseudocode}
\usepackage{algcompatible}
\usepackage{outlines}
\usepackage{orcidlink}
\usepackage{lipsum,comment}
\usepackage{mathtools}
\usepackage{tabularx,flushend}
\usepackage{amsthm}
\newtheorem{remark}{Remark}

\theoremstyle{definition}

\def\BibTeX{{\rm B\kern-.05em{\sc i\kern-.025em b}\kern-.08em
T\kern-.1667em\lower.7ex\hbox{E}\kern-.125emX}}

\newcommand{\trans}[0]{^{\mathsf{T}}}

\newcommand{\herm}[0]{^{\mathsf{H}}}

\newcommand{\Real}[1]{\Re\{{#1}\}}

\newacronym{RPE}{RPE}{radar parameter estimation}
\newacronym{OTFS}{OTFS}{orthogonal time frequency space}
\newacronym{AFDM}{AFDM}{affine frequency division multiplexing}
\newacronym{MIMO}{MIMO}{multiple-input multiple-output}
\newacronym{SISO}{SISO}{single-input single-output}
\newacronym{ISAC}{ISAC}{integrated sensing and communications}
\newacronym{3D}{3D}{three-dimensional}
\newacronym{2D}{2D}{two-dimensional}
\newacronym{1D}{1D}{one-dimensional}
\newacronym{RX}{RX}{receiver}
\newacronym{TX}{TX}{transmitter}
\newacronym{BF}{BF}{beamforming}
\newacronym{mmWave}{mmWave}{millimeter-wave}
\newacronym{SotA}{SotA}{state-of-the-art}
\newacronym{ULA}{ULA}{uniform linear array}
\newacronym{QAM}{QAM}{quadrature amplitude modulation}
\newacronym{ISFFT}{ISFFT}{inverse symplectic finite Fourier transform}
\newacronym{SFFT}{SFFT}{symplectic finite Fourier transform}
\newacronym{AWGN}{AWGN}{additive white Gaussian noise}
\newacronym{OFDM}{OFDM}{orthogonal frequency division multiplexing}
\newacronym{OCDM}{OCDM}{orthogonal chirp division multiplexing}
\newacronym{BS}{BS}{base station}
\newacronym{UE}{UE}{user equipment}
\newacronym{DFT}{DFT}{discrete Fourier transform}
\newacronym{IDFT}{IDFT}{inverse discrete Fourier transform}
\newacronym{TD}{TD}{time-domain}
\newacronym{wlg}{wlg}{without loss of generality}
\newacronym{CP}{CP}{cyclic prefix}
\newacronym{DAFT}{DAFT}{discrete affine Fourier transform}
\newacronym{IDAFT}{IDAFT}{inverse discrete affine Fourier transform}
\newacronym{CPP}{CPP}{\textit{chirp-periodic} prefix}
\newacronym{IDZT}{IDZT}{inverse discrete Zak transform}
\newacronym{DZT}{DZT}{discrete Zak transform}
\newacronym{ICI}{ICI}{inter-carrier interference}
\newacronym{BER}{BER}{bit error rate}
\newacronym{DoF}{DoF}{degrees-of-freedom}
\newacronym{FD}{FD}{full-duplex}
\newacronym{SIMO}{SIMO}{single-input multiple-output}
\newacronym{MISO}{MISO}{multiple-input single-output}
\newacronym{AoD}{AoD}{angle-of-departure}
\newacronym{AoA}{AoA}{angle-of-arrival}
\newacronym{RF}{RF}{radio frequency}
\newacronym{SIM}{SIM}{stacked intelligent metasurfaces}
\newacronym{FIM}{FIM}{flexible intelligent metasurface}
\newacronym{FPGA}{FPGA}{field programmable gate array}
\newacronym{UPA}{UPA}{uniform planar array}
\newacronym{CC}{CC}{communication-centric}
\newacronym{I/O}{I/O}{input-output}
\newacronym{iid}{i.i.d.}{independent and identically distributed}
\newacronym{IoT}{IoT}{internet of things}
\newacronym{V2X}{V2X}{vehicle-to-everything}
\newacronym{NTN}{NTN}{non-terrestrial network}
\newacronym{LEO}{LEO}{low-earth orbit}
\newacronym{THz}{THz}{terahertz}
\newacronym{EM}{EM}{electromagnetic}
\newacronym{STAR-RIS}{STAR-RIS}{simultaneously transmitting and reflecting reconfigurable intelligent surface}
\newacronym{DoA}{DoA}{direction-of-arrival}
\newacronym{DD}{DD}{doubly-dispersive}
\newacronym{ODDM}{ODDM}{orthogonal delay-Doppler division multiplexing}
\newacronym{LoS}{LoS}{line-of-sight}
\newacronym{NLoS}{NLoS}{non-line-of-sight}
\newacronym{6G}{6G}{sixth generation}
\newacronym{MPDD}{MPDD}{metasurfaces-parameterized DD}
\newacronym{GaBP}{GaBP}{Gaussian Belief Propagation}
\newacronym{MSE}{MSE}{mean-squared-error}
\newacronym{sIC}{soft IC}{soft interference cancellation}
\newacronym{soft RG}{soft RG}{soft replica generation}
\newacronym{BG}{BG}{belief generation}
\newacronym{SGA}{SGA}{scalar Gaussian approximation}
\newacronym{CLT}{CLT}{central limit theorem}
\newacronym{PDF}{PDF}{probability density function}
\newacronym{QPSK}{QPSK}{quadrature phase-shift keying}
\newacronym{LMMSE}{LMMSE}{linear minimum mean square error}
\newacronym{SNR}{SNR}{signal-to-noise ratio}
\newacronym{QoS}{QoS}{quality of service}
\newacronym{CoV}{CoV}{calculus of variations}
\newacronym{CAPA}{CAPA}{continuous aperture array}
\newacronym{FCAPA}{FCAPA}{flexible continuous aperture array}
\newacronym{GL}{GL}{Gauss-Legendre}
\newacronym{DDC MIMO}{DDC MIMO}{DD continuous MIMO}
\newacronym{B5G}{B5G}{beyond fifth generation}
\newacronym{VR}{VR}{virtual reality}
\newacronym{XR}{XR}{extended reality}
\newacronym{ITN}{ITN}{intelligent traffic networks}
\newacronym{SAGIN}{SAGIN}{space-air-ground integrated network}
\newacronym{UAV}{UAV}{unmanned aerial vehicle}
\newacronym{MUSIC}{MUSIC}{Multiple Signal Classification}
\newacronym{ICC}{ICC}{integrated communication and computing}
\newacronym{SINR}{SINR}{signal-to-interference-plus-noise ratio}
\newacronym{WSR}{WSR}{weighted sum rate}
\newacronym{ARPU}{ARPU}{average rate per user}
\newacronym{BCD}{BCD}{block coordinate descent}
\newacronym{PDE}{PDE}{partial differential equation}
\newacronym{EL}{EL}{Euler-Lagrange}
\newacronym{TCA}{TCA}{tightly coupled array}
\newacronym{ELAA}{ELAA}{extremely large-aperture arrays}
\newacronym{LIS}{LIS}{large intelligent surface}
\newacronym{CSI}{CSI}{channel state information}
\newacronym{RIS}{RIS}{reconfigurable intelligent surface}
\newacronym{KKT}{KKT}{Karush-Kuhn-Tucker}
\newacronym{MoM}{MoM}{method of moments}
\newacronym{SVD}{SVD}{singular value decomposition}
\newacronym{RWG}{RWG}{Rao-Wilton-Glisson}
\newacronym{EFIE}{EFIE}{electric field integral equation}
\newacronym{rms}{rms}{root mean square}

\begin{document}

\title{A Novel Framework for the Characterization of Continuous Electromagnetic Manifolds\vspace{-0.5ex}}

\author{\IEEEauthorblockN{Kuranage Roche Rayan Ranasinghe,$\!\!^*$ Miguel Rodrigo Castellanos,$\!\!^\dag$ Giuseppe Thadeu Freitas de Abreu$^*$}
\IEEEauthorblockA{$^*$\textit{School of Computer Science and Engineering, Constructor University, Bremen, Germany} \\
$^\dag$\textit{Department of Electrical Engineering and Computer Science, University of Tennessee, Knoxville, USA} \\
Emails: [kranasinghe, gabreu]@constructor.university, mrcastellanos@utk.edu}\vspace{-4ex}}

\maketitle

\begin{abstract}
A unified framework for the characterization of continuous \ac{EM} manifolds for arbitrary \ac{MIMO} system geometries is presented. 
The \ac{EM} manifold refers to the set of all physically realizable radiated field vectors, parameterized by the array excitation, that encodes the full spatial structure of the antenna system including near-field phase variations, polarization, and mutual coupling.
Building upon the discrete moment-matrix formulation, the proposed framework addresses three fundamental limitations simultaneously: (i)~point-source near-field modeling errors in the radiation operator; (ii)~confinement of the beamforming space to the $N$-dimensional subspace dictated by hardware
port count; and (iii)~restriction to linear (\acs{1D}) array geometries.
Each mesh element is modeled as a \ac{2D} planar patch, whose spatially averaged Green's function is evaluated via \acf{GL} quadrature, yielding superior near-field accuracy at negligible additional cost.
A continuous feeding function $w(\mathbf{p})\in L^2(\mathcal{S}_\mathrm{T})$ is introduced as the infinite-dimensional limit of the $N$-port network, enabling optimization over a higher dimensional current subspace, decoupled from hardware constraints.
Full-wave MATLAB Antenna Toolbox validation confirms near-field accuracy improvements over the \ac{SotA} baseline for both linear and planar array geometries, while maintaining reasonable computational complexity. 
\end{abstract}


\glsresetall

\vspace{-1ex}
\section{Introduction}
\vspace{-1ex}

Multi-antenna technologies have been central to the evolution of wireless systems, exploiting an ever-growing number of spatial \acp{DoF}~\cite{PoonTIT2005}. 
For the past two decades, the dominant modeling paradigm -- steering vectors of complex exponentials derived from far-field plane-wave assumptions -- has been adequate because array apertures were small relative to communication distances, mutual coupling was manageable, and beamforming objectives were confined to azimuth/elevation steering.

Three converging trends now collectively invalidate these assumptions and necessitate electromagnetically-consistent treatment. 
\emph{First}, the transition to extremely large aperture arrays~(XL-MIMO) and \ac{CAPA}~\cite{WangTWC2025-CAPAOptimal} dramatically expands the Rayleigh distance $d_R = 2D^2/\lambda$, where $D$ is the aperture diameter. 
For a $1\,\text{m}$ aperture at $5\,\text{GHz}$, $d_R \approx 333\,\text{m}$,
placing typical users firmly in the near field, where spherical wavefront
curvature, spatial non-stationarity, and polarization mixing cannot be
neglected~\cite{wang2025analytical}. 
\emph{Second}, sub-wavelength element spacing introduces significant mutual coupling that the scalar model ignores entirely; the embedded element currents $\mathbf{j}_n(\mathbf{s})$ obtained from a full-wave solver implicitly encode all coupling effects, and any model that bypasses this physics systematically misrepresents the available \acp{DoF}.
\emph{Third}, near-field sensing and \ac{ISAC} applications require precise localization of scatterers at distances where radial Green's function components are non-negligible~\cite{elbir2024near}; errors in the radiation operator translate directly into localization bias.
Together, these trends represent a qualitative shift in operating regime: the far-field scalar model is no longer a mild approximation but a structural mismatch with the underlying physics, motivating the electromagnetically
rigorous framework developed below.

Historically, the foundational \ac{EM} manifold framework for modeling these arrays relied heavily on spatially discrete representations \cite{YangTAP2019,FriedlanderTSP2020,CastellanosTWC2025}. 
These standard models parameterize the array response via a moment matrix derived from a point-source (Dirac delta) approximation of each mesh element.
While conceptually powerful for traditional far-field beamforming, this approach carries three restrictive limitations. 
The point-source radiation operator inherently accumulates systematic approximation errors in near-field modeling due to the rapid spatial variations of the Green's function.
Moreover, the effective beamforming space is artificially confined to an $N$-dimensional current subspace dictated strictly by the physical hardware port count, thereby overlooking the broader potential of the continuous \ac{EM} surface.
Additionally, while the discrete model in~\cite{CastellanosTWC2025} is applicable to arbitrary antenna geometries, its local tangent-frame construction and reported numerical results are developed specifically for linear \ac{1D} array topologies.

To simultaneously address these limitations, it is critical to adopt a continuous \ac{EM} framework. 
Unlike discrete optimization models that face significant computational burdens or rely on suboptimal approximations for continuous functional programming, a continuous \ac{CoV} approach fundamentally decouples the optimization space from hardware constraints. 
By modeling the mesh elements as realistic \ac{2D} planar patches evaluated via Gauss-Legendre quadrature, the radiation operator achieves superior near-field accuracy without incurring a prohibitive computational cost. 
Furthermore, introducing a continuous feeding function as the infinite-dimensional limit of the $N$-port network lifts the discrete beamforming subspace restrictions, maximizing \acp{DoF} and seamlessly supporting the arbitrary planar topologies essential for modern \ac{MIMO} advancements \cite{YuanAWPL2022}.
 
This paper addresses all three limitations within a unified \ac{EM} framework. The specific contributions are:
\begin{itemize}
    \item \textbf{Patch-based radiation operator:} Each mesh element is modeled as a realistic \ac{2D} planar patch; the spatially averaged Green's function is computed via tensor-product \ac{GL} quadrature, yielding consistently more accurate near-field representations at negligible additional cost.
 
    \item \textbf{Continuous feeding framework:} A continuous feeding function $w(\mathbf{p})\!\in\!L^2(\mathcal{S}_\mathrm{T})$ is introduced as the infinite-dimensional limit of the $N$-port network, enabling optimization over a $K$-dimensional current subspace ($K\!\gg\!N$) that is decoupled from hardware constraints.
    The consequent functional optimizations that can be formulated and solved over this function are presented in an extended journal version of this manuscript.

    {
    \item \textbf{Planar geometry support:} The patch-based tangent-frame construction explicitly extends the radiation operator to planar \ac{2D} surface geometries and is numerically validated for both linear and planar arrays, complementing the linear-array results of~\cite{CastellanosTWC2025}.
}
\end{itemize}
 
\textit{Notation:}
Scalars, vectors, and matrices are denoted by plain, bold lowercase,
and bold uppercase letters. $\mathbf{A}\trans$ and $\mathbf{A}\herm$
denote transpose and Hermitian transpose.
{$L^2(\mathcal{S}_T)$ denotes the Hilbert space of square-integrable
complex-valued functions on $\mathcal{S}_T$ with inner product $\langle f, g \rangle_{L^2} = \int_{\mathcal{S}_T} f^*(\mathbf{s})\, g(\mathbf{s})\, \mathrm{d}\mathbf{s}$, over a surface with points characterized by $\mathbf{s}.$}
The Lebesgue measure of $\mathcal{S}$ is $|\mathcal{S}|$.
{$\Real{\cdot}$ denotes the real part of a complex scalar, vector, or matrix.}

\section{State of the Art: Continuous Surface\\with Discrete Feeds}
\label{sec:sota}
 
\subsection{Electric Field Integral and Green's Function}
 
Let $\mathbf{s} = [s_x, s_y, s_z]\trans \in \mathcal{S}_\mathrm{T}$
and $\mathbf{r} = [r_x, r_y, r_z]\trans \in \mathcal{S}_\mathrm{R}$
denote points on the transmit and receive surfaces {$\mathcal{S}_\mathrm{T}$ and $\mathcal{S}_\mathrm{R}$}, respectively,
embedded in a homogeneous medium (e.g., free space).
The electric field at $\mathbf{r}$ produced by a current density
distribution $\bm{j}(\mathbf{s})$ over $\mathcal{S}_\mathrm{T}$ is
given by the radiation integral~\cite{PoonTIT2005}
\begin{equation}
\label{eq:E_integral}
\mathbf{e}(\mathbf{r})
= \int_{\mathcal{S}_\mathrm{T}}
\mathbf{G}(\mathbf{r}, \mathbf{s})\,\bm{j}(\mathbf{s})\,
{\rm d}\mathbf{s} \in \mathbb{C}^{3\times 1},
\end{equation}
where the free-space dyadic Green's function
$\mathbf{G}(\mathbf{r},\mathbf{s})\in\mathbb{C}^{3\times3}$
is~\cite{YuanAWPL2022}
\begin{equation}
\label{eq:G}
\mathbf{G}(\mathbf{r},\mathbf{s})
= \Bigl(\mathbf{I}_3 + \tfrac{\nabla\nabla}{\kappa^2}\Bigr)
\frac{e^{-j\kappa\|\mathbf{r}-\mathbf{s}\|}}
     {4\pi\|\mathbf{r}-\mathbf{s}\|}.
\end{equation}

{
Here $\kappa = 2\pi/\lambda$ is the free-space wavenumber with wavelength $\lambda$, $\mathbf{I}_3 \in \mathbb{R}^{3\times 3}$ is the identity matrix, and $\nabla\nabla$ is the dyadic (outer) gradient operator whose $(i,j)$ entry is $\partial^2/\partial r_i \,\partial r_j$ applied to the scalar Green's function $g(\mathbf{r},\mathbf{s}) = e^{-j\kappa\|\mathbf{r}-\mathbf{s}\|} / (4\pi\|\mathbf{r}-\mathbf{s}\|)$.
The operator $\mathbf{I}_3 + \nabla\nabla/\kappa^2$ encodes both transverse
and longitudinal field components. 
$\mathbf{G}(\mathbf{r},\mathbf{s})$ is smooth for $\mathbf{r} \neq \mathbf{s}$ but singular on the diagonal $\mathbf{r} = \mathbf{s}$; a property that motivates the spatial averaging introduced in Section~\ref{sec:contribution}.
}

Equation \eqref{eq:E_integral} can be used to derive many functional optimization problems aimed at designing $\bm{j}(\mathbf{s})$ in order to achieve specific beamforming/multiplexing goals \cite{WangTWC2025-CAPAMU,WangTWC2025-CAPAOptimal,WangTWC2026-CAPAMIMO}.
 
\subsection{Discrete $N$-Port Feeding and Continuous Steering Matrix}
 
Let $\mathbf{w} = [w_1, \dots, w_N]\trans\in\mathbb{C}^{N\times1}$
denote the weight vector of an $N$-port feed network exciting
$\mathcal{S}_\mathrm{T}$.
The superimposed current density induced by these ports is
\vspace{-1ex}
\begin{equation}
\label{eq:j_disc}
\vspace{-1ex}
\bm{j}(\mathbf{s};\mathbf{w})
= \sum_{n=1}^N w_n\,\bm{j}_n(\mathbf{s}) \in \mathbb{C}^{3\times1},
\end{equation}
where $\bm{j}_n(\mathbf{s})$ is the elementary current density at
surface point $\mathbf{s}$ due to a 1 Ampere {\ac{rms}} excitation at
port~$n$ with all other ports open-circuited.

The matrix $\mathbf{M} \in \mathbb{C}^{3K \times N}$ stacks the sampled
embedded currents: column $n$ is $\mathbf{m}_n = [\mathbf{j}_n(\mathbf{s}_1)\trans, \ldots, \mathbf{j}_n(\mathbf{s}_K)\trans]\trans \in \mathbb{C}^{3K}$, where $\{\mathbf{s}_k\}_{k=1}^K$ are the mesh centroids.
Substituting \eqref{eq:j_disc} into~\eqref{eq:E_integral} and
exchanging the sum and integral yields
\vspace{-1ex}
\begin{equation}
\label{eq:E_disc}
\vspace{-1ex}
\mathbf{e}(\mathbf{r};\mathbf{w})
= \sum_{n=1}^N w_n
\underbrace{
\int_{\mathcal{S}_\mathrm{T}}
\mathbf{G}(\mathbf{r},\mathbf{s})\,\bm{j}_n(\mathbf{s})\,
\mathrm{d}\mathbf{s}
}_{\triangleq\,\mathbf{a}_n(\mathbf{r})\,\in\,\mathbb{C}^{3\times1}}
= \mathbf{A}(\mathbf{r})\,\mathbf{w},
\end{equation}
where the continuous steering matrix is $\mathbf{A}(\mathbf{r})
\triangleq [\mathbf{a}_1(\mathbf{r}),\dots,\mathbf{a}_N(\mathbf{r})]
\in\mathbb{C}^{3\times N}$.
 
Equation~\eqref{eq:E_disc} is the continuous analogue of~\cite[Eq.~(24)]{CastellanosTWC2025}: the radiated field arises from the \emph{full} continuous current distribution over $\mathcal{S}_\mathrm{T}$, not from $N$ Hertzian dipoles; only the excitation mechanism is discrete.
This distinction matters practically: since the actual current on the
surface is physical (governed by the surface boundary conditions), the
steering vectors $\mathbf{a}_n(\mathbf{r})$ carry richer near-field
information than a point-source approximation can recover.

\subsection{Relationship to MoM and RWG Basis Functions}
The \ac{MoM} with \ac{RWG} basis functions~\cite{Makarov2001} is the canonical numerical technique for solving the \ac{EFIE} for unknown surface currents. 
In \ac{MoM}/\ac{RWG}, the surface current is expanded as $\mathbf{j}(\mathbf{s}) = \sum_{k} I_k \boldsymbol{\Lambda}_k(\mathbf{s})$, where $\boldsymbol{\Lambda}_k$ are the \ac{RWG} functions defined on triangular mesh pairs, and the coefficients $\{I_k\}$ are obtained by enforcing boundary conditions via Galerkin testing. 
The resulting impedance matrix encodes mutual couplings through dyadic Green's function integrals, including singular self-terms that require specialized quadrature.

The present framework differs fundamentally in purpose. 
Rather than solving for unknown currents, the goal is to \emph{characterize the
radiated field manifold} for prescribed port excitations, using the physical currents $\mathbf{j}_n(\mathbf{s})$ already determined by a full-wave solver as an input. 
The matrix $\mathbf{M}$ stacks these embedded element currents, extracted by exciting each port individually with all others open-circuited. 
The radiation operator $\mathbf{K}(\mathbf{r})$ then maps these currents to radiated fields at observation points $\mathbf{r} \notin \mathcal{S}_T$, where the integrand is smooth and standard quadrature applies without singularity treatment. 
The contribution here is therefore not an alternative current solver, but a
higher-fidelity \emph{post-processing} operator that replaces the zeroth-order centroid rule of~\cite{CastellanosTWC2025} with spatially averaged patch operators, improving near-field accuracy without revisiting the underlying \ac{MoM} solve.

A related body of work characterizes antenna arrays directly from their
far-field patterns~\cite{PoonTIT2005} or via scalar channel models that
absorb mutual coupling into a fixed correction matrix~\cite{Wallace2004}. 
While computationally convenient, these approaches conflate the physical current distribution with its far-field projection, discarding near-field phase structure. 
The discrete dipole~(Hertzian dipole) approximation of~\cite{CastellanosTWC2025} partially recovers near-field information by retaining the full dyadic Green's function, but applies it at mesh centroids only; a zeroth-order rule whose error grows as the observation point approaches the surface.

\vspace{-1ex}
\section{Proposed Framework: Continuous Feeding\\and Patch-Based Radiation}
\label{sec:contribution}

\subsection{Continuous Feeding Function}
\vspace{-1ex}
 
The limitation to an $N$-dimensional beamforming subspace in \eqref{eq:E_disc} is intrinsic to any finite-port architecture.
To overcome it, we consider the limit $N\to\infty$, replacing the discrete superposition in~\eqref{eq:j_disc} with
\vspace{-1ex}
\begin{equation}
\label{eq:j_cont}
\vspace{-1ex}
\bm{j}(\mathbf{s})
= \int_{\mathcal{S}_\mathrm{T}}
w(\mathbf{p})\,\bm{j}(\mathbf{s},\mathbf{p})\,{\rm d}\mathbf{p},
\end{equation}
where $\mathbf{j}(\mathbf{s},\mathbf{p}) \in \mathbb{C}^{3\times 1}$ is the embedded element current density at $\mathbf{s}$ due to a unit feed excitation at $\mathbf{p}$.
This motivation is also backed by \ac{SotA} on optimal designs for the construction of such feeds as done in \cite{GustafssonTAP2013,JelinekTAP2017}.

\vspace{-1ex}
\begin{remark}[Discrete-to-continuous convergence]
\label{rem:convergence}
The integral in~\eqref{eq:j_cont} is the formal $L^2$ limit of the
discrete superposition in~\eqref{eq:j_disc}.
For a sequence of uniform port grids with spacing $\Delta p\to 0$ and
weights $w_n = w(\mathbf{p}_n)\Delta p$, the Riemann sums converge in
$L^2(\mathcal{S}_\mathrm{T})$ to~\eqref{eq:j_cont}, provided
$w\in L^2(\mathcal{S}_\mathrm{T})$ and $\bm{j}(\mathbf{s},\mathbf{p})$
is square-integrable jointly in $(\mathbf{s},\mathbf{p})$.
\end{remark}

\vspace{-2ex}
\begin{remark}[Physical realizability of the continuous feed]
\label{rem:feed_realizability}
The continuous feeding model~\eqref{eq:j_cont} is presented as a theoretical construct defining an upper bound on achievable beamforming performance (see Remark~\ref{rem:bound}). 
Its physical realization is, however, non-trivial. 
In standard antenna modeling, each feed port imposes a localized boundary condition, typically a delta-gap voltage source or a coaxial probe, that uniquely determines the current distribution on the entire surface through the governing integral equation.
Under such a model, specifying $w(\mathbf{p})$ as a continuous function over
$\mathcal{S}_T$ would over-constrain the system: the surface current is already determined by the full-wave boundary value problem once the discrete port excitations are fixed, and $w(\mathbf{p})$ cannot be prescribed independently. 
The transition from~\eqref{eq:j_disc} to~\eqref{eq:j_cont} is therefore best understood as a mathematical limiting argument (Remark~\ref{rem:convergence}) rather than a prescription for a physically realizable feed architecture. 
The consequent functional optimizations, and the design of structured $N$-port networks whose aggregate response approximates a target $w(\mathbf{p})$, are deferred to the extended journal version.
\end{remark}
\vspace{-1ex}
 
Substituting \eqref{eq:j_cont} into~\eqref{eq:E_integral} and exchanging the order of integration yields the fundamental continuous field equation
\vspace{-1ex}
\begin{equation}
\label{eq:E_cont}
\vspace{-1ex}
\mathbf{e}(\mathbf{r})
= \int_{\mathcal{S}_\mathrm{T}} w(\mathbf{p})\,
\underbrace{
\int_{\mathcal{S}_\mathrm{T}}
\mathbf{G}(\mathbf{r},\mathbf{s})\,\bm{j}(\mathbf{s},\mathbf{p})\,
\mathrm{d}\mathbf{s}
}_{\triangleq\,\mathbf{a}(\mathbf{r},\mathbf{p})\,\in\,\mathbb{C}^{3\times1}}
\mathrm{d}\mathbf{p},
\end{equation}
where $\mathbf{a}(\mathbf{r},\mathbf{p})$ is the \emph{continuous steering vector}; the field at $\mathbf{r}$ due to a unit feed at $\mathbf{p}$, and the continuous analogue of $\mathbf{a}_n(\mathbf{r})$.

\vspace{-1.5ex}
\begin{remark}[Aperture bound interpretation]
\label{rem:bound}
The continuous model~\eqref{eq:E_cont} represents a theoretical upper bound on beamforming performance achievable from $\mathcal{S}_\mathrm{T}$.
Any physical $N$-port system confines the current distribution to an $N$-dimensional subspace of $L^2(\mathcal{S}_\mathrm{T})$; the continuous model subsumes all $N$-port realizations and provides a principled, hardware-agnostic benchmark.
\end{remark}
\vspace{-1.5ex}
 
For a receiver with polarization direction $\mathbf{u}_r\in\mathbb{R}^{3\times1}$, the scalar received field is
\vspace{-1ex}
\begin{equation}
\label{eq:e_scalar}
\vspace{-1ex}
e(\mathbf{r})
= \mathbf{u}_r\herm\mathbf{e}(\mathbf{r})
= \int_{\mathcal{S}_\mathrm{T}}
w(\mathbf{p})\,a(\mathbf{r},\mathbf{p})\,{\rm d}\mathbf{p},
\end{equation}
where $a(\mathbf{r},\mathbf{p})
\triangleq \mathbf{u}_r\herm\mathbf{a}(\mathbf{r},\mathbf{p})
\in\mathbb{C}$
is the scalar continuous steering function, which inherits
$L^2(\mathcal{S}_\mathrm{T})$ square-integrability from
$\bm{j}(\mathbf{s},\mathbf{p})$ for any $\mathbf{r}\notin
\mathcal{S}_\mathrm{T}$.

\vspace{-1ex}
\subsection{Patch-Based Radiation Operator}
\label{subsec:patch}
\vspace{-1ex}
 
The second contribution is a higher-order radiation operator that replaces the zeroth-order centroid rule of~\eqref{eq:Kk_ps}.
Each of the $K$ mesh elements is modeled as a \ac{2D} planar patch of effective size $L_k = \sqrt{A_k}$ rather than a point source.
A local tangent frame is constructed for each patch: the dominant current direction
\vspace{-1ex}
\begin{equation}
\label{eq:d1}
\vspace{-1ex}
\hat{\mathbf{d}}_{k,1}
= \frac{\sum_{n=1}^N \Real{\mathbf{M}_{(k)}\,\mathbf{e}_n}}
       {\bigl\|\sum_{n=1}^N \Real{\mathbf{M}_{(k)}\,\mathbf{e}_n}
        \bigr\|}.
\end{equation}

{
Here $\mathbf{e}_n \in \mathbb{R}^N$ is the $n$-th standard basis vector, $\mathbf{M}_{(k)} \triangleq \mathbf{M}(3k{-}2\,:\,3k,\,:) \in \mathbb{C}^{3 \times N}$ are the three rows of $\mathbf{M}$ corresponding to segment $k$, and $\hat{\mathbf{d}}_{k,2}$ is the unit vector in the tangent plane of $\mathcal{S}_T$ at $\mathbf{s}_k$ satisfying $\hat{\mathbf{d}}_{k,2} \perp \hat{\mathbf{d}}_{k,1}$ and $\hat{\mathbf{d}}_{k,2} \perp \hat{\mathbf{n}}_k$, where $\hat{\mathbf{n}}_k$ is the outward surface normal at $\mathbf{s}_k$.
}
If the mean current vanishes (e.g., for symmetric geometries), $\hat{\mathbf{d}}_{k,1}$ is taken as the dominant left singular vector of $\Real{\mathbf{M}_{(k)}}$.
 
The spatially averaged Green's function over patch~$k$ is then
\vspace{-1ex}
\begin{equation}
\label{eq:Kk}
\vspace{-1ex}
\mathbf{K}_k(\mathbf{r})
\!=\! \frac{1}{4}\int_{-1}^1\!\int_{-1}^1
\!\!\!\!\mathbf{G}\!\left(
\mathbf{r},\,\mathbf{s}_k
+ \tfrac{L_k}{2}\bigl(\xi\hat{\mathbf{d}}_{k,1}
+ \eta\hat{\mathbf{d}}_{k,2}\bigr)\!
\right)\!
d\xi\,d\eta,
\end{equation}
where $L_k = \sqrt{A_k}$ is the effective patch side length, the factor $1/4$ is the Jacobian of the coordinate change $(x,y) \in [-L_k/2, L_k/2]^2 \to (\xi,\eta) \in [-1,1]^2$ normalized by the patch area $A_k = L_k^2$, yielding a spatially averaged (per unit current moment) field operator.

Since $\mathbf{r}\notin\mathcal{S}_\mathrm{T}$, the integrand is smooth in $(\xi,\eta)$, and the \ac{2D} integral is evaluated efficiently via \ac{GL} quadrature, given by
\vspace{-1ex}
\begin{align}
\label{eq:GL}
\vspace{-1ex}
\mathbf{K}_k(\mathbf{r}) &\approx \\
&\hspace{-5ex}\frac{1}{4} \sum_{q_1=1}^{N_q} \sum_{q_2=1}^{N_q} \omega_{q_1} \omega_{q_2} \, \mathbf{G}\left(\mathbf{r}, \mathbf{s}_k + \frac{L_k}{2} \big(\xi_{q_1} \hat{\mathbf{d}}_{k,1} + \eta_{q_2} \hat{\mathbf{d}}_{k,2}\big)\right), \nonumber
\end{align}
where $\{\xi_q,\omega_q\}_{q=1}^{N_q}$ are the \ac{GL} nodes and weights on $[-1,1]$.

Because the Green's function is smooth away from its source, \ac{GL} quadrature converges exponentially fast in $N_q$.
Empirically, for sub-wavelength elements ($A_k\leq\lambda^2/100$), $N_q = 2$ (four integration points per patch) is sufficient to achieve relative errors below $10^{-3}$ for all observation points.
Stacking the patch operators, the discrete-port radiated field is
\begin{equation}
\label{eq:e_disc_patch}
e(\mathbf{r})
= \mathbf{u}_r\herm\mathbf{K}(\mathbf{r})\mathbf{M}\,\mathbf{w},
\end{equation}
where $\mathbf{K}(\mathbf{r}) = [\mathbf{K}_1(\mathbf{r}), \ldots,
\mathbf{K}_K(\mathbf{r})] \in \mathbb{C}^{3 \times 3K}$ is the full patch
radiation matrix, $\mathbf{M} \in \mathbb{C}^{3K \times N}$ is the embedded
current matrix defined above, and $\mathbf{u}_r \in \mathbb{R}^3$ is the
unit polarization vector of the receive antenna.

\vspace{-1ex}
\subsection{Numerical Approximation in the SotA}
\label{subsec:sota_approx}
 
For numerical evaluation, \cite{CastellanosTWC2025} partitions $\mathcal{S}_\mathrm{T}$ into $K$ mesh segments with centroids
$\{\mathbf{s}_k\}_{k=1}^K$, areas $\{A_k\}_{k=1}^K$, and
stacks the embedded current vectors into the matrix
$\mathbf{M}\in\mathbb{C}^{3K\times N}$.
The radiated field contribution of segment~$k$ is then approximated
by evaluating the Green's function at the centroid alone:
\vspace{-1ex}
\begin{equation}
\label{eq:Kk_ps}
\vspace{-1ex}
\mathbf{K}_k^\mathrm{ps}(\mathbf{r})
= \mathbf{G}(\mathbf{r},\mathbf{s}_k),
\end{equation}
which is equivalent to a zeroth-order (midpoint) quadrature rule for $\int_{A_k}\mathbf{G}(\mathbf{r},\mathbf{s})\,\mathrm{d}\mathbf{s}$.

While this approximation is accurate in the far-field -- where $\mathbf{G}$ varies slowly over a sub-wavelength element -- it accumulates systematic error in the near-field, where the rapid spatial variation of $\mathbf{G}$ demands higher-order integration.
 
\section{Numerical Evaluation}
\label{sec:results}
 
\subsection{Setup}
 
The proposed model is validated against full-wave \ac{EM} simulations performed using the MATLAB Antenna Toolbox (EHfields function) at $5$~GHz.
Two element types and two array topologies are considered: \textit{(i)} half-wave dipole elements and \textit{(ii)} bowtie triangular elements, each configured as an \textit{8-element linear array} and a \textit{$2\times4$-element planar array}, and each evaluated at two element spacings ($\lambda/2$ and $4\lambda$), yielding twelve total configurations.
{The planar array is oriented in the $yz$-plane with its main radiation direction along the $(+)x$-axis.}
For each configuration, the embedded current matrix $\mathbf{M}\in\mathbb{C}^{3K\times N}$ is extracted from MATLAB by exciting each port in isolation and recording the \ac{MoM} segment currents, giving $K = 1{,}152$ segments and $N = 16$ ports.
All results are reported at azimuth $120^\circ$ and elevation $30^\circ$ {for ease of exposition.}
The relative error metric is
\vspace{-1ex}
\begin{equation}
\label{eq:err}
\vspace{-1ex}
\text{Relative\ error}
= \frac{\|\mathbf{e}_\mathrm{sim}(\mathbf{r}) - \mathbf{e}(\mathbf{r})\|}
       {\|\mathbf{e}_\mathrm{sim}(\mathbf{r})\|},
\end{equation}
where $\mathbf{e}_\mathrm{sim}(\mathbf{r})$ is the full-wave simulated field and $\mathbf{e}(\mathbf{r})$ is the model prediction.
The baseline is the point-source model of~\cite{CastellanosTWC2025} which evaluates $\mathbf{G}(\mathbf{r},\mathbf{s}_k)$ at each centroid without spatial averaging, as described in~\eqref{eq:Kk_ps}.
 
\subsection{Field Accuracy: Linear Arrays}

\begin{figure}[H]
    \centering
    \begin{subfigure}{\columnwidth}
        \centering
        \includegraphics[width=\columnwidth]{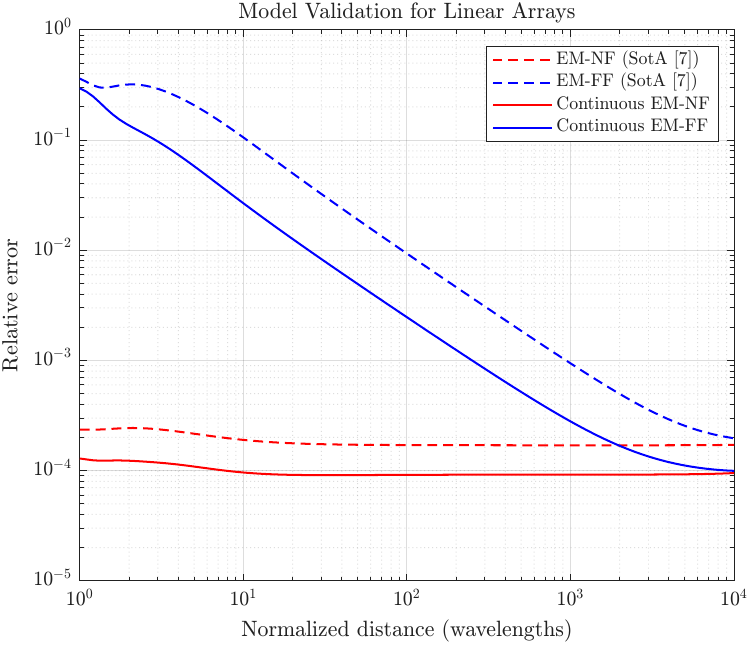}
        \vspace{-4ex}
        \caption{8-element dipole array, $\lambda/2$ spacing.}
        \label{fig:lin_dip_half}
    \end{subfigure}
    \\[0.5ex]
    \begin{subfigure}{\columnwidth}
        \centering
        \includegraphics[width=\columnwidth]{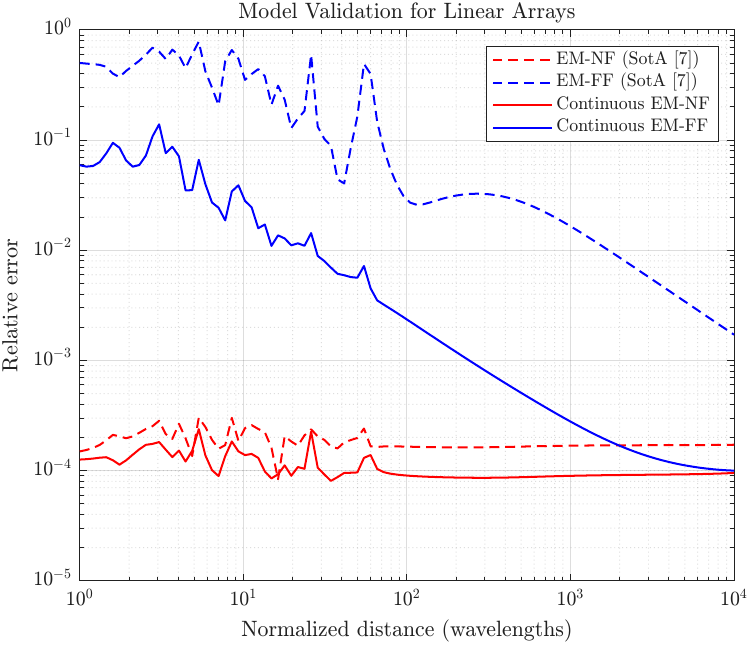}
        \vspace{-4ex}
        \caption{8-element dipole array, $4\lambda$ spacing.}
        \label{fig:lin_dip_4lam}
    \end{subfigure}
    \vspace{-4ex}
    \caption{Relative field error vs. distance for linear arrays.}
    \label{fig:valid_linear}
    \vspace{-2ex}
\end{figure}

Fig.~\ref{fig:valid_linear} shows the relative field error as a function of distance for three linear array configurations.
The proposed patch-based model consistently achieves lower error than the \ac{SotA} point-source baseline across the entire range, from the reactive near-field to the Fraunhofer far-field.
In the near-field region (distances of order $\lambda$ to a few $\lambda$), where the rapid spatial variation of the Green's function renders centroid evaluation inadequate, the accuracy gain of the proposed operator is most pronounced.
As distance increases and the field smooths, both models converge to the same far-field limit, as expected.
This confirms that the proposed framework is strictly more accurate in the near field without any regression in the far field.

For the $\lambda/2$ case (Fig.~\ref{fig:lin_dip_half}), the relative error of the proposed model falls well below that of the baseline throughout. 
Increasing the element spacing to $4\lambda$ (Fig.~\ref{fig:lin_dip_4lam}) broadens the angular spread of the embedded current patterns and leads to higher absolute errors for both models, but the proposed method maintains its consistent advantage. 
Configurations using bowtie triangular elements confirm the same accuracy across all distance regimes, with the proposed patch-based operator consistently outperforming the \ac{SotA}. 
Results are omitted for brevity; the patch-based formulation requires no changes across element geometries since $\hat{\mathbf{d}}_{k,1}$ in~\eqref{eq:d1} adapts to the dominant current direction of each element type.

\vspace{-1ex}
\subsection{Field Accuracy: Planar Arrays}
\vspace{-1ex}

\begin{figure}[H]
    \centering
    \includegraphics[width=\columnwidth]{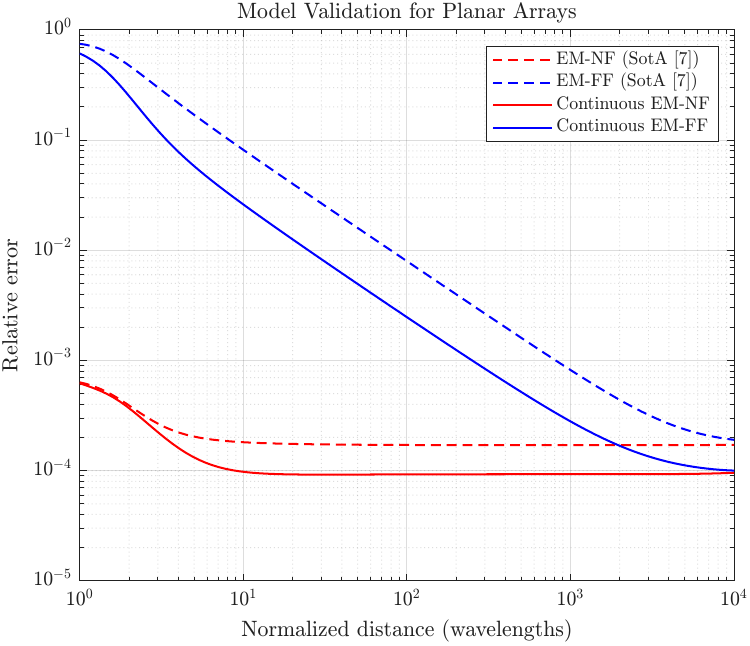}
    \vspace{-4ex}
    \caption{Relative field error vs. distance for a $2\times4$ dipole array planar array with $\lambda/2$ spacing.}
    \label{fig:valid_planar}
    \includegraphics[width=\columnwidth]{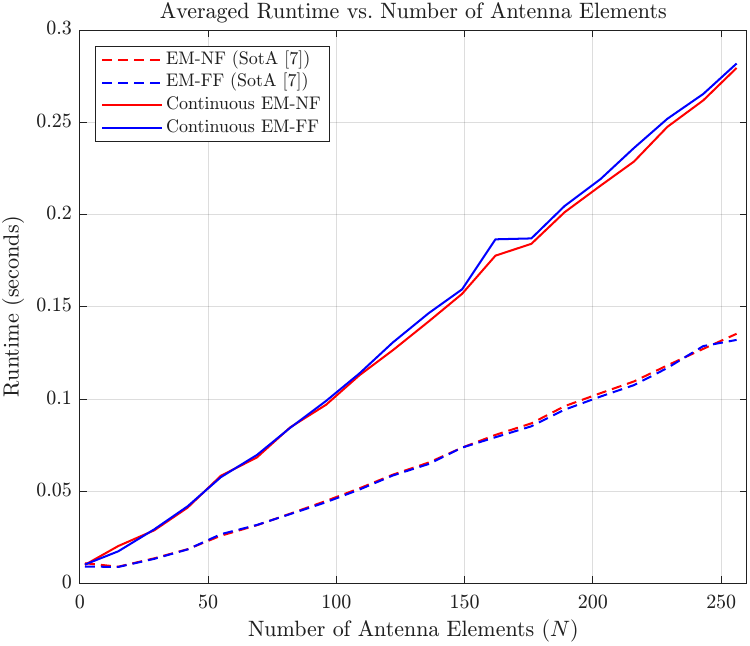}
    \vspace{-3ex}
    \caption{Average runtime comparison: \ac{SotA} vs.\ proposed model.}
    \label{fig:runtime}
    \vspace{-3ex}
\end{figure}

Fig.~\ref{fig:valid_planar} presents the same evaluation for the planar array configurations.
The accuracy gains of the proposed model are fully preserved in the transition from linear to planar geometries, a regime not explicitly addressed in~\cite{CastellanosTWC2025}.

This validates the extension to arbitrary \ac{2D} surface geometries enabled by the patch-based local tangent-frame construction of~\eqref{eq:Kk}--\eqref{eq:GL}: the dominant current directions $\hat{\mathbf{d}}_{k,1}$ adapt automatically to the planar topology of $\mathcal{S}_\mathrm{T}$, with no modification to the radiation operator.

For $4\lambda$ spacing, both models exhibit higher absolute errors due to the broader angular spread of the embedded current patterns, but the proposed model maintains its consistent advantage over the \ac{SotA} in the near-field region.
This trend is identical to the linear-array case (Fig.~\ref{fig:lin_dip_4lam}) and the corresponding planar result is therefore omitted.

\vspace{-1ex}
\subsection{Computational Complexity}
\vspace{-1ex}

Fig.~\ref{fig:runtime} compares the average computation time of the two implementations as a function of the number of antenna elements $N$.
The \ac{SotA}~\cite{CastellanosTWC2025}) applies the zeroth-order centroid rule but evaluates the dyadic Green's function directly, bypassing the rotation-matrix transforms.
The proposed method evaluates four Green's function calls per patch as in~\eqref{eq:GL}.
Comparing the results isolates the true accuracy -- complexity trade-off of the proposed higher-order operator: the $N_q = 2$ \ac{GL} scheme introduces a constant factor of four relative to the zeroth-order baseline, independent of $N$, while delivering the near-field accuracy gains demonstrated in Figs.~\ref{fig:valid_linear}--\ref{fig:valid_planar}.

\section{Conclusion}
\label{sec:conclusion}
 
We have presented a unified framework for the characterization of continuous \ac{EM} manifolds for \ac{MIMO} systems, addressing near-field accuracy, hardware-decoupled \acp{DoF}, and planar geometry support simultaneously.
A patch-based \ac{GL} radiation operator replaces the point-source centroid approximation of~\cite{CastellanosTWC2025}, delivering consistent near-field accuracy improvements across all array configurations tested -- linear and planar, dipole and bowtie, half-wave and wide-spaced -- with similar runtime cost.
A continuous feeding function $w(\mathbf{p})\in L^2(\mathcal{S}_\mathrm{T})$
lifts the effective beamforming space from the $N$-dimensional port subspace, and makes consequent optimizations hardware-agnostic.
Future work will address the design of structured $N$-port feed networks that approach the continuous bound.

\bibliographystyle{IEEEtran}
\bibliography{references}

\end{document}